\documentstyle[twocolumn,aps,psfig,here]{revtex}

\begin{document}

\twocolumn[\hsize\textwidth\columnwidth\hsize\csname 
              @twocolumnfalse\endcsname



\title{A stringy black string \\and a stringy black hole in four dimensions}


\author{Bj{\o}rn Jensen
 and Svend E. Hjelmeland
}

\address{Institute of Physics, University of Oslo,\\
P.O. Box 1048, N-0316 Blindern, Oslo 3, Norway}



\bibliographystyle{unsrt}

\maketitle

 
\begin{abstract}
We uplift the static three dimensional black hole solution found by
Banados, Teitelboim and Zanelli (BTZ) into four dimensional space time.
In this way we obtain a black string solution with a relativistic string source,
as well as a new black hole solution which is also generated by a
relativistic ``stringy'' source. It is shown that when passing continuously
from the region of the one dimensional parameter spaces
which characterize these solutions, containing naked
singular strings or naked singular ``points'', and into the region containing 
black strings or black holes, the metrics ``blow up'' at a critical point 
in both parameter spaces. We show that a similar ``separation'' 
mechanism can be introduced in three dimensions.
In this way we also obtain a generalization of the BTZ solution.
The ``separation'' mechanism, and its immediate consequences,
offers an hitherto missing technical argument needed in order
 to exclude the naked singularities (the ``mass gap'') from the space of ``physically permissible'' solutions in three dimensional Einstein theory coupled to a negative cosmological constant. {\em PACS 
nos}.: {\bf 04.40.Nr , 04.20.Cv}
\end{abstract}

\vskip2pc]



 
It is somewhat surprising that a 
black hole solution exists in three dimensional (3D) Einstein gravity
coupled to a {\em negative} cosmological constant \cite{BT} (the BTZ solution).
This solution has received an immense attention since it was first published. 
The possibility to construct a black hole in 3D, even though localized sources
do not curve their exterior in 3D, can (in some sense) be seen to be
coupled to the existence of an angle
deficit outside any gravitating ``object'' in 3D. Such a deficit typically
induces a ``conical'' geometry in spatial sections. Hence, even
though the Riemann curvature in vacuum is unaffected
by the presence of sources,
the angle deficit will induce a ``residual''
gravitational effect, such that light rays passing by a 
source are deflected.
A singular point source in 3D anti de Sitter (AdS) space will give rise to
a black hole depending on the amount of residual focusing which is
induced by the source. 
When $M$ is a ``natural'' mass measure,
a point source in 3D AdS space gives rise to a black hole when
$M>0$, and a naked singularity when $-1<8GM<0$. 
This last interval was identified as a ``mass gap'' in \cite{BT}.
$8GM=-1$ corresponds to the pure AdS vacuum, and $M=0$ is identified
with the black hole ``ground state''.
The static BTZ solution is locally and in ``polar'' coordinates explicitly given by
\begin{equation}
ds^2=-Bdt'^2+B^{-1}dr'^2+r'^2d\theta'^2
\end{equation}
where 
$B=-8GM +|\, ^{(3)}\Lambda |r'^2$,
with $M$ representing the ``mass'' of the black hole,
and $^{(3)}\Lambda$ is the 3D cosmological constant.
In the fundamental
work \cite{BT} eq.(1) was prescribed to hold for $r'>0$. In this work we will let
$r'\geq 0$.
The range of the angular coordinate
$\theta '$ will be specified later. 
However, we note that in pure AdS vacuum we must have $0\leq\theta '\leq 2\pi$.
We will in the following
set the gravitational coupling constant in 4D to
unity, while we will keep the three dimensional analog $G$ explicitly in 3D.
Greek indices will in general refer to 4D, $(a,b,...)$ to 3D while $(m,n,...)$
are reserved for 2D quantities.  

In \cite{BT} the mass gap was excluded from the ``physical spectrum''.
This of course is naturally motivated since naked singularities are,
for various reasons,
troublesome in general. However, in \cite{BT} this exclusion of the mass gap
from the space of physically permissible solutions was done ``by hand''.
In fact, it is nothing in the equations themselves which compel
us to accept such a conclusion. 
The primary objective for the present study is to cast some light upon this issue. 
We do this by first uplifting the static  
BTZ solution into 4D. We
obtain a black string solution,
and a new spherical symmetric black hole solution
in this way. Each solution is fully characterized by a one dimensional parameter space
(at a fixed cosmological constant). Even though these solutions are very different
globally, 
they display
a common behavior near critical points ${\cal P}$ in the associated parameter spaces,
corresponding to the point $M=0$ in the BTZ solution. In both cases it is impossible
to go continuously
from the black string/hole sector, and into the sector exhibiting naked singular
strings or naked singular ``points''. At the points ${\cal P}$ the metrics blow up. 
In the case of a single string source some of the coordinates used are further forced into
the complex plane. The black hole geometry changes orientation when the
critical point is crossed.
 Hence, in 4D the black string/hole sectors is effectively separated
from the naked singularity sectors.
It is then shown that a mechanism identical to the one at work in the 4D black hole
configuration can be
implemented in 3D. Our approach provides one analytical avenue 
which can be followed in order to exclude
the naked singularities from the ``physical spectrum'' also in 3D. 
The rest of our presentation 
is organized as follows.
We next present, and to some extent discuss, the stringy black string
and stringy black hole solutions. We then ``implement'' the
naked singularity-black hole ``separation'' mechanism in 3D. 
 

We will first describe in some detail our general considerations which are
important in our derivations of
new black structures in 4D. Our philosophy is rather
straight forward. We simply ask what kind of structures in 4D, with the
extra spatial dimension being an isometrical direction, allows for the BTZ solution
to be found as an hyper surface; that is, a test particle moving in such a surface
should not be ``able'' to tell whether it is moving in a true 3D 
BTZ space or in 4D.
Since our investigation is primerely aimed at casting
some light on the mass gap issue, we
will also demand that the mechanism for the existence of 
new black structures in 4D is
essentially the same as in 3D. Let us now endow 4D space time with a slicing
defined by a space like vector field
$\vec{n}$, which is orthonormal to each surface in the slicing. Let us also
denote the coordinate in the new spatial dimension in 4D by $\lambda$, such that each
hyper surface $\Sigma_\lambda$ in the slicing is characterized by $\vec{n}(\lambda )$.
$\lambda$ may take any value in $\Re$,
or in a compact sub set of $\Re$. We parameterize
each $\Sigma_\lambda$ surface with coordinates $\{ x^c\}=\{ t,r,\theta \}$.
Let the geometry in each $\Sigma_\lambda$ be denoted by $h_{ab}$.
If $h_{a'b'}$ is the geometry in 3D then
a simple embedding of the BTZ solution 
into any $\Sigma_\lambda$ is given by
$\Sigma :\,  \{x^{c'}\}\rightarrow \{x^c=\alpha^cx^{c'}\}\,\, ,\,\,
\Sigma (h_{a'b'})=h_{ab}$
where we allowed for three {\em arbitrary} and non zero constants
$\{\alpha^t,\alpha^r,\alpha^\theta\}\equiv \{\alpha^c\}$
in the coordinatization in 4D. We will fix these constants in our
explicit constructions later in this paper. Since
we have two everywhere commuting Killing vectors in the 3D geometry 
$(\vec{\xi}=\partial_{t'}\, ,\, \vec{\eta}=\partial_{\theta '})$,
and since $\Sigma$ maps $h_{a'b'}$ {\em isometrically} into $h_{ab}$ 
the image of these vectors will also be everywhere commuting
Killing vectors
in the 4D geometry.  Clearly, since $\vec{n}$ is taken to be
hyper surface orthonormal we have the following simple isometry (sub) algebra in 4D :
$[\vec{\xi},\vec{\eta}]=
[\vec{\xi},\vec{n}]=[\vec{\eta},\vec{n}]=0$.
The 4D Ricci scalar $^{(4)}R$ can be written 
\begin{equation}
2\, ^{(4)}R=2\, ^{(3)}R(h)+h^{-1}(2\pi^{ab}\pi_{ab}-\pi^2)
\end{equation}
where
$\pi^{ab}=\sqrt{h}(K^{ab}-Kh^{ab})\, ,\, K_{ab}= n_{a;b}$, $K=K^a\, _a$,
$^{(3)}R$ denotes the 3D Ricci scalar, and $h_{ab}=g_{ab}-n_an_b$
the induced metric in $\Sigma_\lambda$.
It is easy to see that the following relation holds true
\begin{equation}
2\pi^{ab}\dot{h}_{ab}=2Nh^{-1/2}(2\pi^{ab}\pi_{ab}-\pi^2)
\end{equation}
Since $\vec{n}$ is assumed to be a Killing vector we have
 $\dot{h}_{ab}=0\,\, (\Rightarrow {\cal L}_{\vec{n}}g_{ab}
= 0)$,  
and $\pi^{ab}\pi_{ab}-\frac{1}{2}\pi^2=0$. Hence, 
when confined to a fixed $\Sigma_\lambda$ hyper surface we get back
the Einstein-Hilbert action in 3D. This is a necessary but certainly, as
we will see later, not
a sufficient condition to be met in order to get the BTZ solution in
$\Sigma_\lambda$. Let us now turn to some explicit constructions.

{\bf Stringy black string:}
We will first consider possibly the simplest, and in some sense most natural,
uplifting of the BTZ solution,
namely when the uplifted structure can be interpreted as a relativistic string.
It is well known that there exists an
intimate relation between the geometries induced by a singular relativistic
 string in
4D, and the geometry induced by a point particle in 3D.
 We will also demand as a
further ``natural'' constraint, that when the cosmic fluid is set
to vanish in the uplifted geometry, the geometry induced by such
a string is regained.
The geometry of a singular relativistic string is given by \cite{Jensen1,Jensen2}
\begin{equation}
ds^2=\frac{1}{(1-4\mu )}(-dt^2+dr^2+d\lambda^2+(1-4\mu )^2r^2d\theta^2)
\end{equation}
where $\mu$ is a measure of the energy per unit length along the string.
It is assumed that $0\leq\theta\leq 2\pi$. 
This geometry reflects the defining property of such a string source,
namely ``Lorentz'' invariance along the string axis \cite{Jensen2}.
The line element eq.(4) is clearly singular at $\mu =\frac{1}{4}$, and it changes
signature when $\mu >\frac{1}{4}$. This behavior will be the key to the
naked string - black string separation mechanism as we will see below.
We may ``get rid'' of
the ``conformal'' factor in this expression by passing to other coordinates
defined by
$T=(1-4\mu )^{-1/2}t\, ,\, R=(1-4\mu )^{-1/2}r$ and 
$Z=(1-4\mu )^{-1/2}\lambda$, such
that the geometry can be written in the more familiar form
\begin{equation}
ds^2=-dT^2+dR^2+dZ^2+(1-4\mu )^2R^2d\theta^2
\end{equation}
Relative to these coordinates $\mu$ is the ADM energy density in the string
relative to Minkowski space time \cite{Jensen2}.
The cosmic ``fluid'' given by
\begin{eqnarray}
8\pi T^\mu\, _\nu =|\, ^{(4)}\Lambda |\delta^\mu\, _\nu +2|\, ^{(4)}\Lambda |
\delta^\mu\, _Z\delta^Z\, _\nu\, ,
\end{eqnarray}
which describes an effective 4D (anisotropic) cosmological constant, allows the
following geometry to be induced
\begin{eqnarray}
ds^2=-Ud t'^2+U^{-1}dr'^2 +r'^2d\theta'^2 +dZ^2
\end{eqnarray}
$U=\tilde{\alpha}+|\, ^{(4)}\Lambda |r'^2$, and $\tilde{\alpha}$ is some
integration constant. When $^{(4)}\Lambda$ is set to 
vanish we want to get back a relativistic string.
This constraint implies that $\tilde{\alpha}=1-4\mu$, $t'=\tilde{\alpha}^{-1}t$,
$r'=r$, $\theta '=\tilde{\alpha}^{1/2}\theta$.
These relations fully specify the transformation $\Sigma$ above, and in particular
the range of $\theta '$.
The resulting geometry can then be written in the form
\begin{eqnarray}
ds^2&=&\frac{1}{\tilde{\alpha}}(-(\tilde{\alpha} +|\, ^{(4)}\Lambda|r^2)\frac{dt^2}{\tilde{\alpha}}+\nonumber\\
&+&\frac{\tilde{\alpha}dr^2}{(\tilde{\alpha} +|\, ^{(4)}\Lambda |r^2)}
+\tilde{\alpha}^2r^2d\theta^2+d\lambda^2)
\end{eqnarray}
We may also identify $|\, ^{(4)}\Lambda |=|\, ^{(3)}\Lambda |$.
It is clear that eq.(8) also can be derived by a direct integration of the
field equations such that we would have $\{\alpha^c\}=\{1\}$.
This point will be illustrated when we come to the stringy black hole below.
Comparing eq.(1) with eq.(7) we find that the mass parameter
$M$ in eq.(1) can be written in terms of $\mu$ as $8GM=(4\mu -1)$. 
The above line element describes a naked singular relativistic (cosmic)
string embedded into AdS space. This line element has the correct
signature provided $\mu <\frac{1}{4}$. 
However, the line element is clearly singular at $\mu =\frac{1}{4}$, 
and it displays a 
serious signature problem when $\mu >\frac{1}{4}$. In order to have at least a formally
acceptable geometry in the $\mu >\frac{1}{4}$ sector with signature $(-+++)$ we
must analytically continue the geometry  
through the assignments
$\theta\rightarrow i\theta$
and $\lambda\rightarrow i\lambda$. 
In these ``complex'' coordinates the metric  
describes a physically acceptable
black string for all values $\mu >\frac{1}{4}$. Clearly, at $\mu =\frac{1}{4}$
the geometry still remains singular.
This is an interesting demonstration of an
``absolute'' separation of the solution space into a naked singularity sector and
a black string sector.

{\bf Stringy black hole:}
A natural question is whether it is possible to uplift the BTZ solution in such
a way that we get a spherically symmetric black hole.
Since our aim
is to understand some features in 3D it is natural to ask whether it is
possible to find such a solution using relativistic strings.
Surprisingly, this is not a difficult task. One possibility is the following 
four-geometry
\begin{equation}
ds^2=-(q+\frac{|\, ^{(4)}\Lambda |r^2}{3})\frac{dt^2}{\kappa^2}+
\frac{dr^2}{(q+\frac{|\, ^{(4)}\Lambda |r^2}{3})}
+r^2d\Omega^2_2
\end{equation}
where $d\Omega^2_2=d\theta^2+\sin^2\theta d\lambda^2$
is the {\em regular} infinitesimal surface area on $S^2$, i.e. $\lambda\in [0,\pi ]$,
$0\leq\theta\leq 2\pi$,
$\kappa$ and $q$ are two independent
integration constants and $\{\alpha^c\}=\{ 1\}$. 
Clearly, $8GM=-q$.
The source for this geometry is, in addition
to the usual (isotropic) cosmological constant, the ``fluid''
$\rho\equiv T^{t}\, _{t}=T^{r}\, _{r}=\frac{(q-1)}{8\pi r^2}$ 
and we may identify $3|\, ^{(3)}\Lambda |=|\, ^{(4)}\Lambda |$.
This source has the same boost invariant structure as the singular relativistic string
source
we considered earlier \cite{Jensen2}. Also note that $\rho\rightarrow 0$ 
as $r\rightarrow\infty$, such
that we asymptotically get back to the pure AdS vacuum. The canonical mass measure
$\tilde{M}$ for this source is given via
$d\tilde{M}=\frac{1}{2} (1-q)dr\equiv\tilde{\mu} dr$. $\tilde{\mu}/(4\pi )$ 
is therefore a natural measure
of the {\em constant} 
mass per unit length along a radially directed ray, again in direct analogy with
the singular string. 
Also note that this source does not break the weak energy condition provided
that $q\leq 1$.
We will demand that when the cosmological constant is set to
vanish the Lorentz invariance of the source is reflected in the induced geometry.
This constraint is also fundamental in order to construct the geometry induced
by a string \cite{Jensen2}. This
constraint fixes $\kappa^2 =q^{-2}=(1-2\tilde{\mu})^{-2}$. 
Hence, when passing from the $\tilde{\mu} <\frac{1}{2}$ sector and into the 
$\tilde{\mu} >\frac{1}{2}$
sector the metric blows up at $\tilde{\mu}=\frac{1}{2}$.
This is very similar to what happend in the 
single string case,
although the singular behavior is less severe in this case since no
coordinate is forced into the complex plane.
Due to the close analogy between this source and 
the properties displayed by a relativistic string,
we will look upon this source as the continuum limit of a system of radially
directed relativistic strings piercing through a common origin.
We will at the end of the paper discuss the character of the singular behavour
of the geometry in some more detail.

{\bf A mechanism in 3D:}
In this communication we have so far constructed two structures with very
different global geometric symmetries, but in which one finds the BTZ solution 
as sub spaces. We have shown that when both of these systems were
considered constructed from relativistic strings, and that a typical
(defining) {\em geometric} symmetry displayed by such strings also
is encoded in the induced geometries, the sectors
of the solutions presented exhibiting event horizons, are separated from the
sectors exhibiting naked singularities. However, our results are in 4D. 
Is there a similar mechanism in 3D ? The essential mechanism in 4D is not
coupled to the presence of a 
cosmological constant. We are then led to focus on the geometry
induced by a single point source in 3D {\em without} a cosmological constant.
Before we explicitly discuss the nature of the geometry induced by a point
particle, we will first find it useful to reformulate the BTZ geometry in terms
of a hamiltonian mass measure.
The physical (ADM) hamiltonian is given by \cite{Hawking}
\begin{eqnarray}
H_{p}=\int_{S_{t'}}H-
\frac{1}{8{\pi}G}
\int_{S_{t'}^{\infty}}[\Delta (NK)-N^{a}\pi_{ab}N^{b}] \label{Ham}
\end{eqnarray}
where $\Delta (NK)\equiv NK-NK_0$ and $H=(N{\cal H}+N^{a}{\cal H}_{a})$.
The hamiltonian of the reference background 
(represented through $N_0K_0$)
has been taken into account in this expression. 
The hamiltonian constraint ${\cal H}$ 
and momentum constraints ${\cal H}_{a}$,
which stem from the variation of the lapse
function $N$ and shift functions $N^a$
respectively are, as usual, given by
${\cal H}=\, ^{(3)}R(h_{ab})-2\, ^{(2)}R(h_{mn})=0$, 
${\cal H}^{n}= D_m(h^{-1/2}\pi^{mn})=0$.
$D$ represents the projection of the three dimensional covariant derivative
into $S_{t'}$, and $h_{mn}$ is the induced geometry in this hyper surface.
The total (ADM) energy $E$ for a solution associated 
with a general time translation is then given by
\begin{eqnarray}
E=-\frac{1}{8{\pi}G}\int_{S_{t'}^{\infty}}d\theta '
\sqrt{g_{\theta '\theta '}}[\Delta (NK)-
N^{a}\pi_{ab}N^{b}] \label{energy}
\end{eqnarray}
To match the BTZ geometry with the reference
geometry, which we take to be ``empty'' AdS space
(hence, the $\theta '$ integration in eq.(11) must 
extend from zero to $2\pi$),
 we consider a large circle at $r'=r'_{0}$.  
Calculating the trace of the extrinsic curvatures at $r'=r'_{0}$ and setting $N^a =0$ yields $4GE=8GM+1$.
$E$ is independent of $\Lambda$ as it should be, since AdS is 
the reference geometry. Clearly, the relation between $M$ and $E$, and the
corresponding relation between $M$ and $\mu$ is exact. 
Letting $\Lambda\rightarrow 0$
when $4GE<1$ in eq.(1) we derive a metric induced by a singularity at the origin.
$E$ is now the ADM mass of the source relative to Minkowski space.
In 4D and dealing with
isolated sources which obey the energy conditions some integration constants can be
determined by demanding that the asymptotic structure is {\em exactly}
Minkowski. In 3D this is impossible since the conical deficit angle extends to
space like infinity. However, we may nevertheless demand, as a very natural ``constraint'',
that some key features of the ``asymptotic Minkowski constraint'' 
in 4D also applies to localized sources
in 3D. One key ingredient in this condition is that the asymptotic geometry is
``boost invariant'' in the radial direction, i.e. photons moving along
a $d\theta '=0$ ray is made to be moving such that $dt'/dr'=\pm 1$ in
the canonical global coordinate system. The point mass geometry derived from the
BTZ solution does not explicitly exhibit this property. However, the geometry
eq.(1) is not the most general solution of the field equations which
give rise to a black hole. It is straightforward to see that the BTZ solution can be 
generalized to
\begin{eqnarray}
ds^2=-f_0^2 Bdt'^2+B^{-1}dr'^2+r'^2d\theta'^2
\end{eqnarray}
where $f_0$ is an independent integration constant. In \cite{BT} this constant
has apparently been set to unity ``by hand''. Our approach is that
{\em all} integration constants should be determined from 
physically motivated boundary conditions
set on the the geometry. The additional boundary condition which we will impose on
the geometry is the ``asymptotic Minkowski constraint'' discussed above.
This fixes $f_0^{-2}=(1-4GE)^2$. Clearly, this induces a similar kind of
singular behavior of the geometry as we saw present in our
earlier constructions. We also observe that the ADM mass relative to eq.(12) 
is $E(1-4GE)^{-1}$. Interestingly, relative to eq.(4) the ADM energy density
in the string is $\mu (1-4\mu )^{-1}$ (which is the ``dressed'' tension
discussed in \cite{Jensen2}).
These conclusions can also be reached by another, 
and perhapse a somewhat more ``naive'' approach.
Starting from the metric ansatz
$ds^2=-e^{2\alpha}dt'^2+e^{2\beta}dr'^2+r'^2d\theta'^2$,
and assuming that $0\leq\theta '\leq 2\pi$,
it follows from the vacuum field equations that $\alpha$ and $\beta$ must
equal two {\em independent} constants $\alpha_0$ and $\beta_0$. 
We could have included an extra $r'$-dependent function in $g_{\theta '\theta '}$.
Our choice is dictated by the form of the angular part of the BTZ solution eq.(1).
Since the ADM measure above is defined relative to AdS space, it is natural
to define mass measures relative to Minkowski space when $^{(3)}\Lambda =0$. Hence,
when we prescribe the existence of a point mass at the origin,
in the coordinate system above, we characterize it
(rather ``naively'') by the distribution
$T_{t't'}=m\delta (r')/(4\pi r')$ where $\delta$ is the usual Dirac
delta function in flat space. From the field equations 
we then find that $e^{-2\beta_0}=C(1-4 Gm/C)$ where
$C$ is an integration constant.
We can compute a gravitating energy $E_g$ for this distribution from
the conserved current $j^\mu =T^{\mu\nu}\xi_\nu$, for some appropriately chosen
{\em normalized} Killing vector field $\vec{\xi}$. When we choose to measure the
energy relative to the canonical observer $\vec{\xi}=e^{-\alpha_0}\partial_{t'}$,
we find that $E_g=\int \vec{j}\cdot\vec{n}\sqrt{^{(2)}g}dr'd\theta ' =me^{\beta_0-\alpha_0}$
where we have fixed the normal vector
$\vec{n}=\partial_{t'}$, since (again) the reference space
is taken to be Minkowski space. 
The ``asymptotic Minkowski constraint'' fixes $\alpha_0 =\beta_0$.
We also demand that we get back flat Minkowski
space when $M=0$ something which determines $C=1$. 
This implies that $E_g=m=E$. We may define $m\equiv m_0 e^{-2\beta_0}$ where
$m_0$ is a scalar. When we substitute the expression for $m$ into
$T_{t't'}$ we find that $m_0/(4\pi )$ can be taken to represent a
``covariant'' mass measure of the source.   
It is evident from the above that the point particle geometry 
changes its causal character when $1<4Gm$. In 
order to obtain a ``well behaved'' metric in this sector
we can formally
let $x^{c'}\rightarrow ix^{c'}$.
In these coordinates the above analytically continued geometry offers one
physically acceptable way to describe
space outside a particle with $Gm>\frac{1}{4}$.
We also observe that 
a discrete transformation then exists, which maps
a black hole with mass $m$ in 3D AdS space into a naked conical singularity
in 3D ``complexified'' dS space 
(or visa versa) given by
$x^{c'}\rightarrow ix^{c'}$,
$|^{(3)}\Lambda |\rightarrow -\, |^{(3)}\Lambda |$, $(1-4Gm)\rightarrow -(1-4Gm)$.
The transformed geometry solves the field equations. We have assumed
that $Gm<1/2$. The last restriction on the parameter interval 
is necessary in order to ensure that
no angle surplus arises outside the black hole. In that case an extra naked
singularity will also be found outside the black hole \cite{Steif}. 
The mass $M$ of the particle in dS space, which follows
from the transformation above, is
given via $2GM=1-2Gm$.
It is unclear whether we
should insist on the same causal behavior in the two sectors $4Gm<1$, $4Gm>1$,
in the limit of vanishing cosmological
constant, or not. Clearly, if we do not, the transformation above simply reduces to
$|^{(3)}\Lambda |\rightarrow -\, |^{(3)}\Lambda |$, $(1-4Gm)\rightarrow -(1-4Gm)$.
Also, a conical singularity does no longer appear in the point particle
geometry when $4Gm>1$,
even though an angle deficit effect is
still present.

Clearly,
no curvature invariants are singular at the
critical points of the geometric structures which we have derived in this work.
However, the determinants  of the metric
tensors ($g$) are.
 We further observe that $\sqrt{-g}$ {\em changes sign} in 3D, and in the 4D
spherically symmetric configuration at these points.
Hence, the 
naked singularity 
sectors carry another {\em orientation} than the black hole sectors. 
Clearly, the string in 4D has another kind of singularity structure.
We conclude that in 3D
the black hole sector cannot
be deformed into the naked singularity sector
when a physically well motivated boundary condition on the
black hole geometry is imposed,
 by any orientation preserving
homeomorphism. The black hole ``ground state'' can thus be taken
to be a true vacuum state of the theory as argued in \cite{BT}.
 
{\bf Acknowledgments:} We thank N. Kaloper and J.P.S. Lemos
 for bringing some related previous work
\cite{Kaloper} to our attention.



 \end{document}